\documentclass[aps,prl,superscriptaddress,showpacs,floatfix]{revtex4}

\usepackage{amsmath}
\usepackage{bm}
\usepackage{dcolumn}
\usepackage{graphicx}
\usepackage{subfigure}
\usepackage{url}
\usepackage{wrapfig}
\usepackage{titlesec}

\usepackage{hyperref}
\usepackage{breakurl}

\usepackage{amssymb}
\usepackage{indentfirst}    
\usepackage{float}      
\usepackage{afterpage}  
\usepackage{ulem}
\usepackage{color}

\newcommand{\gevsq}{GeV$^2$}

\newcommand{\qsq}{$Q^2$}

\newcommand{\GEp} {$ {G_{_{_E}} ^{\it p}}$}
\newcommand{\GMp}{$ {G_{_{_M}}^{\it p}}$}
\newcommand{\GEn} {$ {G_{_{_E}} ^{\it n}}$}
\newcommand{\GMn}{$ {G_{_{_M}}^{\it n}}$}
\newcommand{\GD}{$ {G_{_{Dipole}}}$}
\newcommand{\Fonep}{\mbox{${ {F_1^{\it p}}}$}}
\newcommand{\Ftwop}{\mbox{${ {F_2^{\it p}}}$}}

\newcommand{\Foneu}{\mbox{${ {F_1^{\it u}}}$}}

\newcommand{\Foned}{\mbox{${ {F_1^{\it d}}}$}}
\newcommand{\Ftwod}{\mbox{${ {F_2^{\it d}}}$}}
\newcommand{\Fsp}{\mbox{${ {F_{\it s}^{\it p}}}$}}
\newcommand{\PR}{{ Phys. Rev. }}
\newcommand{\PRL}{{ Phys. Rev. Lett. }}

\newcommand{\etal}{{\em et al.}}

\begin{document}

\title{\Large{Flavor Decomposition of Nucleon Form Factors}}

\author{Bogdan Wojtsekhowski}

\address{Thomas Jefferson National Accelerator Facility \\
Newport News, Virginia 23606 USA}

\date{\today}

\begin{abstract}
The nucleon form factors provide fundamental knowledge about the strong interaction.
We review the flavor composition of the nucleon form factors
and focus on an analysis of the possible impact of the $s$-quark contribution.
A future experiment is presented to measure the strange form factor at large momentum transfer.
\end{abstract}

\pacs{PACS numbers:14.20.Dh, 13.40.Gp}

\maketitle

\section{\large{High \qsq~nucleon form factor experiments}}

Nucleon structure investigation using high energy electron scattering has been a successful 
field where many discoveries have been made since the 1955 observation of the proton size~\cite{RH1955}.
The status of current knowledge of the nucleon electromagnetic form factors is reviewed in Ref.~\cite{VP2015}. 
To a large extent, this success is due to the dominance of the one-photon exchange mechanism of
electron scattering as proposed in the original theory~\cite{MR1950}.

The most decisive studies of the partonic structure of the nucleon could be performed when the dominant part of 
the wave function is a 3-quark Fock state. 
This requires large momentum transfer, $Q^2 > 1$ GeV$^2$, when the contribution of the pion cloud is suppressed.
By the early 90s, the data sets at large \qsq~for the proton and the neutron have been found to be in agreement in
the Dipole fit,  \GD$\,=\, (1+Q^2/0.71 [GeV^2])^{-2}$, see Ref.~\cite{PB1995}. 
The SLAC experimental data~\cite{RA1986} on the proton Dirac form factor $F_1^p$ at \qsq~above 10~\gevsq~
have been found to be in fair agreement with the scaling prediction~\cite{LB1979} based on perturbative QCD: \Fonep$\, 
\propto Q^{-4}$,  where \qsq~is the negative four-momentum transfer squared.

New development began with a precision experiment~\cite{CP1989} which made a very productive realization
of a double polarization method suggested in Refs.~\cite{AA1958}. 
The double polarization method has large sensitivity to the typically small electric form factor due to the interference 
nature of the double polarization asymmetry.
It is also less insensitive to the two-photon exchange contributions, which complicates the Rosenbluth method. 

\begin{figure}[ht]
\unitlength 1cm
\begin{minipage}[th]{7cm}
   \centering
   \includegraphics[trim = 5mm 10mm 20mm 5mm, width=1.0\textwidth] {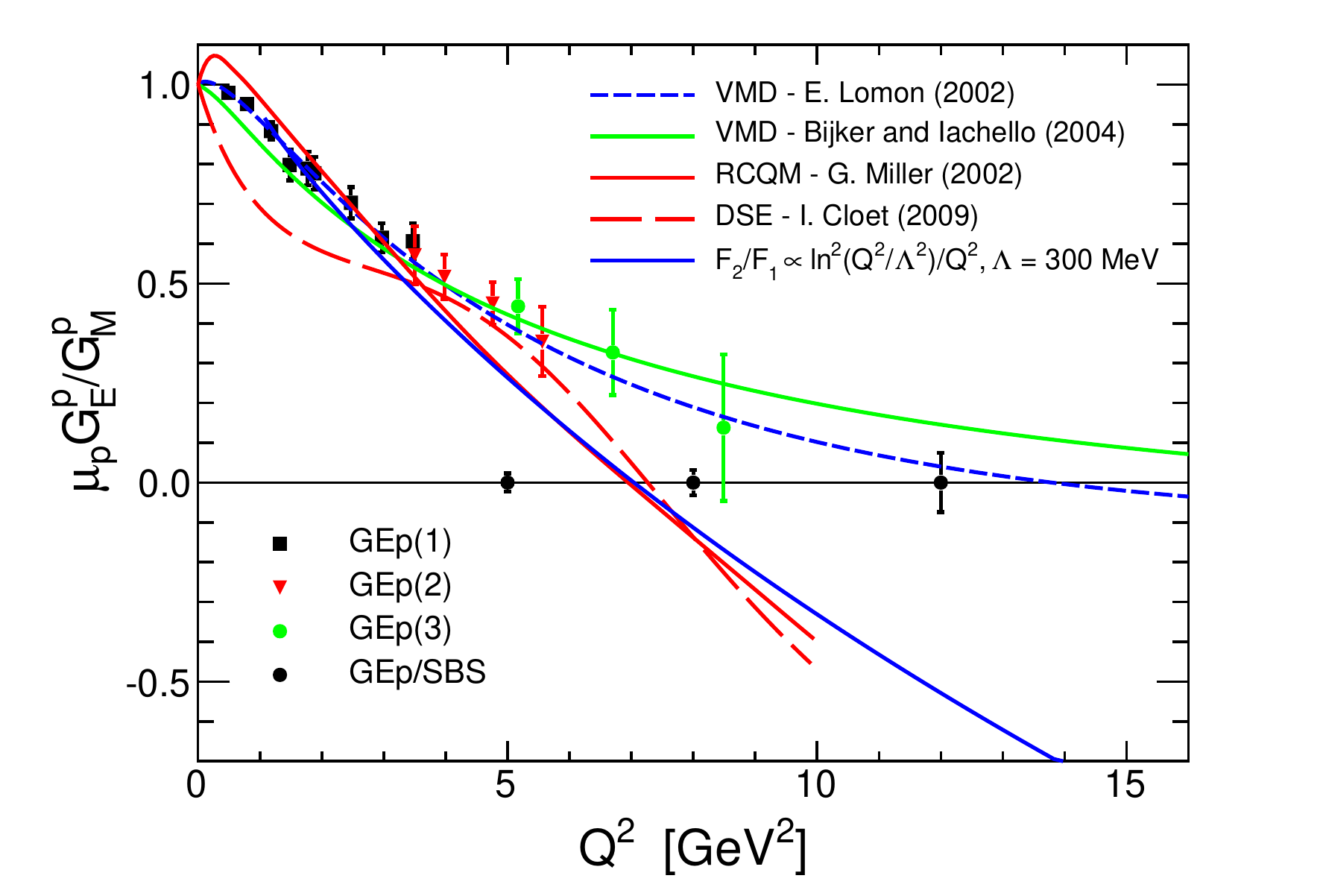} 
\end{minipage}
\hfill
\begin{minipage}[th]{7cm}
\includegraphics[trim = 20mm 15mm 35mm 35mm,width=1.0\textwidth] {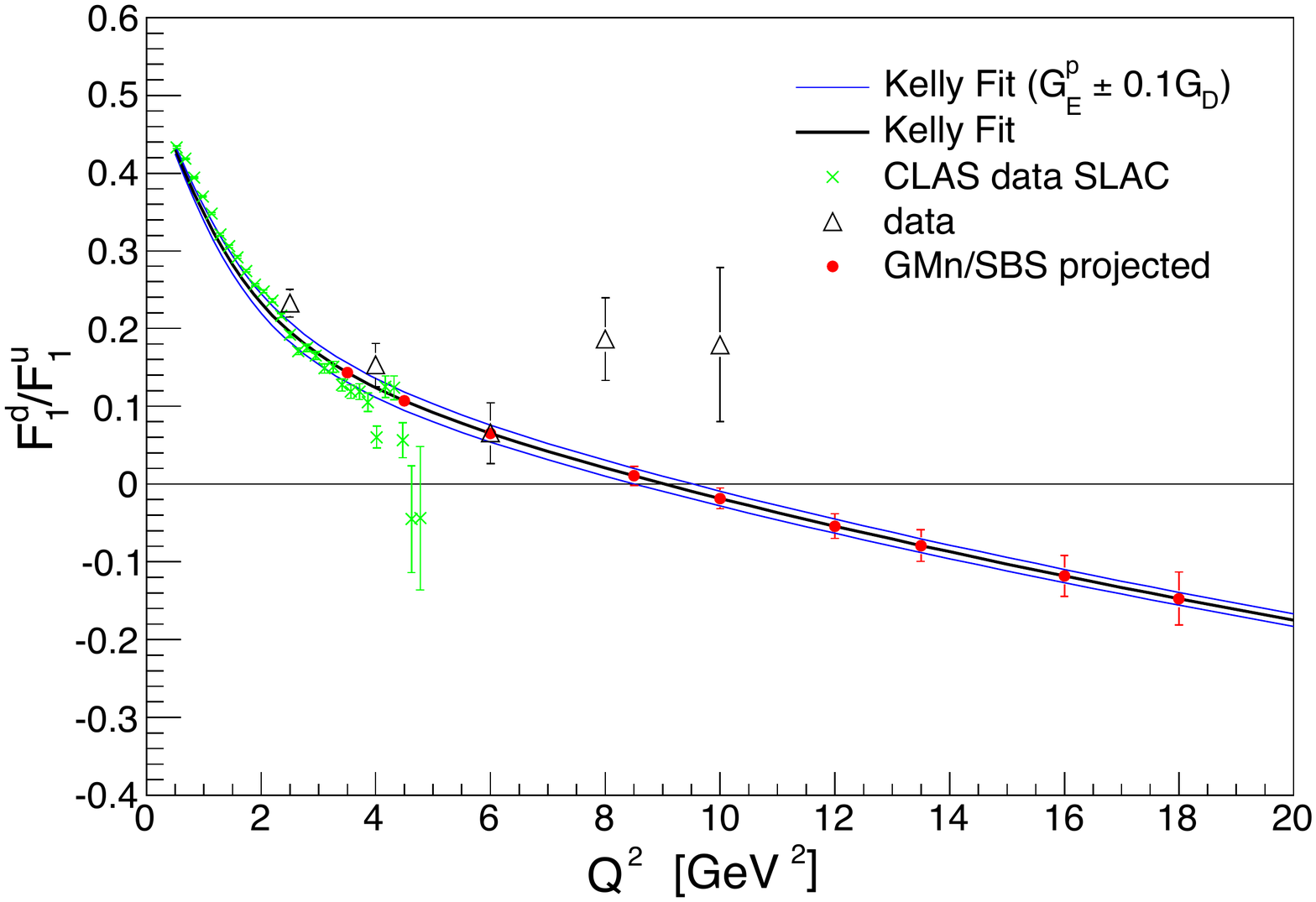}
\end{minipage}
\hskip .5 in
\hfill
   \caption{ Left: Existing data and projected data accuracy for the ratio of the $\mu _p$\GEp/\GMp. 
Right: Ratio of the $d$- and $u$-quark contributions to the proton form factor \Fonep~from
measurement of the \GMn/\GMp (see more in the text). } 
   \label{fig:GEp}
\end{figure}

The experimental results from Jefferson Laboratory~\cite{MJ2000} are shown in Fig.~\ref{fig:GEp} (left). 
The ratio of the proton Pauli form factor \Ftwop\ and the Dirac form factor \Fonep\ have been found to be in disagreement 
with the scaling law $F_2^p/F_1^p \propto 1/Q^2$ (which requires the \GEp~to be proportional to \GMp~for large 
momentum transfer, $\tau >>1 $) suggested in Ref.~\cite{LB1979}.
 
The data for $\mu_p$\GEp/\GMp~revealed an unexpected reduction with \qsq, which also means that 
\Fonep~and \qsq$\times$\Ftwop~for the proton have different $Q^2$ dependencies.
The origin of the scaling prediction violation has been attributed to an effect of the quark orbital momentum 
(so-called ``logarithmic scaling") which provides a very efficient fit of the data for a proton in a wide range 
of the momentum transfer above 1~\gevsq~\cite{BJY2003}.

The measurement of the proton to the neutron cross section ratio in the quasi-elastic nucleon knockout from the deuteron 
was used in JLab's precision experiment of the neutron magnetic form factor for \qsq~up to 4~\gevsq~\cite{JL2009}.
With the latest JLab experiment on the neutron electric form factor~\cite{SR2010}, the data on all four
nucleon form factors have become available in the \qsq~region of 3-quark dominance. 

\begin{figure}[ht]
\unitlength 1cm
\begin{minipage}[th]{7cm}
   \includegraphics[trim = 5mm 10mm 20mm 5mm, width=0.70\textwidth] {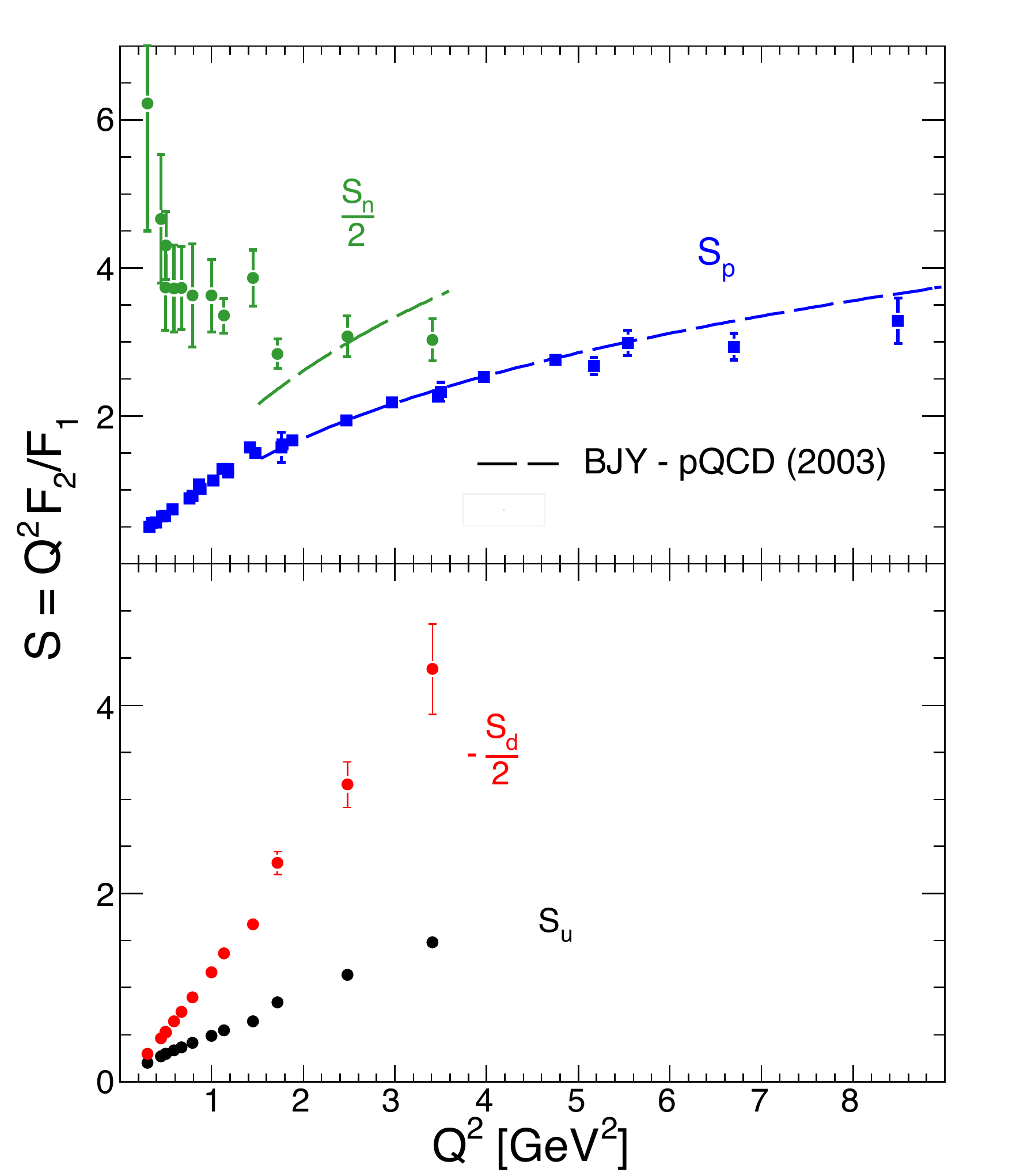} 
\end{minipage}
\hskip .5 in
\begin{minipage}[th]{7cm}
\includegraphics[trim = 20mm 40mm 40mm 27mm, width=0.72\textwidth] {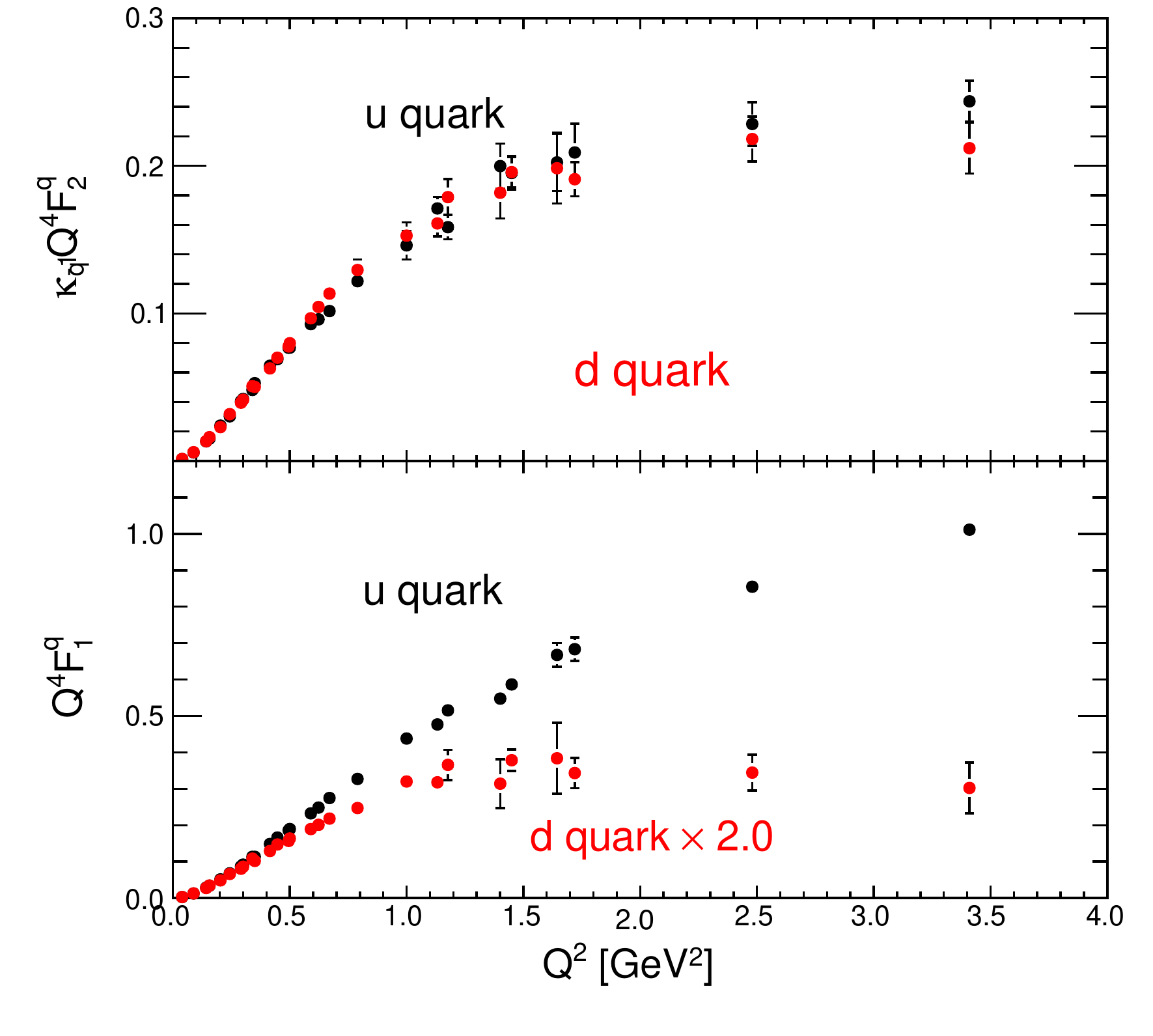}
\end{minipage}
\vskip 0.1 in
     \caption{ The flavor decomposition of proton form factors per Refs.~\cite{CJRW2011,MD2013}.} 
   \label{fig:F12ud}
\end{figure}

The first analysis of these new data for the flavor contributions to the nucleon form factors was reported in Ref.~\cite{CJRW2011}.
The $Q^2F_2/F_1$ for individual flavors as a function of \qsq~shown in Fig.~\ref{fig:F12ud} (left)
does not have any sign of the saturation expected in the case of approaching the pQCD regime.
Analysis shows a large unexpected reduction in the relative size of the $d$-quark contribution to the \Ftwop~form factor, 
which drops by a factor of 3 when \qsq~increases from 1 to 3.4~\gevsq.
A similar result was found in an advanced analysis~\cite{MD2013} with the GPD-based fits of the form factors, see Fig.~\ref{fig:F12ud} (right).

The 0bserved reduction of the $d$-quark contribution to \Ftwop~naturally explains the JLab result for the momentum 
dependence of \Ftwop/\Fonep~without the effect of the quark orbital momentum (at least at \qsq~below 3.4~\gevsq).
The origin of the observed \Foned/\Foneu~reduction with the increase of \qsq~is a subject of significant interest as it could
be the most direct evidence of the di-quark correlations in a nucleon as proposed in Ref.~\cite{CR2007}. 

The flavor decomposition leads to two simple conclusions:
\begin{itemize}
\item The contributions of the $u$-quarks and $d$-quark to the magnetic and electric form 
factors of the proton all have different \qsq~dependencies. 
\item The contribution of the $d$-quark to the \Fonep~form factor at \qsq=3.4~\gevsq~
is three times less than the contribution of the $u$-quarks (corrected for the number of quarks and 
their charge).
\end{itemize}

The second observation suggests that the probability of proton survival after the absorption of a massive virtual
photon is much higher when the photon interacts with a $u$-quark, which is doubly represented in the proton. 
This may be interpreted as an indication of an important role of the $u$-$u$ correlation.
It is well known that the correlation usually enhances the high momentum component and the interaction cross section.
The relatively weak $d$-quark contribution to the \Fonep~indicates a suppression of the $u$-$d$ correlation or a mutual cancellation of different types of $u$-$d$ correlations.

\section{\large{The SBS nucleon form factor program}}

A set of experiments was proposed with the Super BigBite Spectrometer whose large angular acceptance allows
us to advance very significantly the measurements of the \GEp, \GMn, and \GEn~(see Table~\ref{tab:T1}).

\begin{table}[htb]
{\begin{tabular}{@{}cccc@{}} \toprule
Form factor & Reference & \qsq~range, \gevsq & $\Delta G/$\GD (stat/syst) at max \qsq \\ \colrule
\\
\GEp   & \cite{GEpE} &    5-12       &   0.08 / 0.02   \\
\GEn   & \cite{GEnE}  &   1.5-10.2  &  0.23 / 0.07      \\
\GMn   & \cite{GMnE} &   3.5-13.5  &  0.06 / 0.03      \\ 
\\
\botrule
\end{tabular} 
\caption{Upcoming measurements of the nucleon form factors in JLab Hall A with SBS (approved experiments). 
Projected range of \qsq~and accuracy relative to the Dipole form factor at max. value of \qsq.}
\label{tab:T1}}
\end{table}
The first measurement for the neutron magnetic form factor (\GMn) is under preparation for data taking in 2021. 
Fig.~\ref{fig:GEp} (right) shows the projected accuracy for the ratio \Foneu/\Foned~obtained from 
\GMn/\GMp~with systematic uncertainties dominated by the uncertainty of the \GEp/\GMp~ratio.

\section{\large{New experiment for measurement of the strangeness form factor at high \qsq}}

\begin{figure}[ht]
      \includegraphics[width=0.5\textwidth] {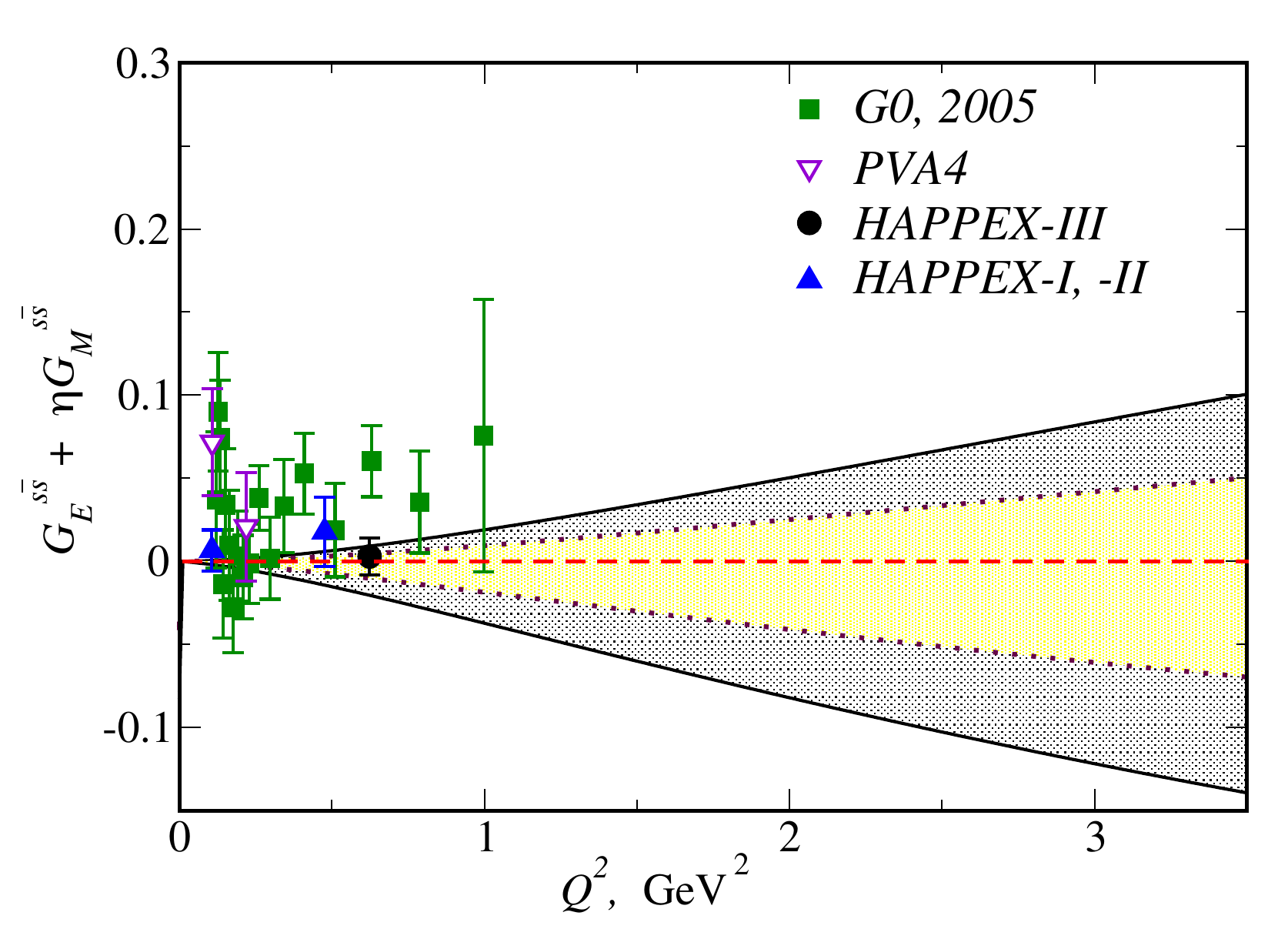} 
     \caption{ The strange form factor vs. momentum transfer data and projections per Refs.~\cite{TH2015,TH2019}.} 
     \label{fig:TH}
\end{figure}

In this section we present the physics motivation and specific ideas for a new experiment for the measurement 
of the \Fsp~by using SBS equipment.
In the original flavor decomposition study~\cite{CJRW2011} we decided to omit the heavier quark contribution motivated 
by the fact that all experimental data on the strangeness form factor of a proton \Fsp~are consistent with zero~\cite{DB2001,DA2012} (in agreement with the lattice calculations).
However, all known experiments were performed for \qsq~below 1~\gevsq.
At the same time, the relative role of the $s \bar s$ in the elastic electron-nucleon scattering  
could be higher at the momentum transfer of 3~\gevsq~\cite{TH2015}.
The recent analysis of the possible value for the strange form factor performed  by T.~Hobbs, M.~Alberg, and J.~Miller
suggests that \Fsp~could be as high as a \GD~(which is 0.03 at Q$^2$=3.4~\gevsq) or even larger, see Fig.~\ref{fig:TH}  from Refs.~\cite{TH2015,TH2019}.

In the one-photon exchange approximation, the amplitude for electron-nucleon elastic scattering can
be written as $ {M^{\rm^{nuc}} = -(4\pi\alpha/Q^2)l^{\mu}\,J_{\mu}^{\rm^{nuc}}}$,
where $\alpha$ is the fine structure constant,  
$l^{\mu}=\overline{e} \gamma^{\mu} e$ is the leptonic vector current, and 
\begin{equation}
J_{\mu}^{\rm^{nuc}} = \langle p (n) |({\textstyle{2\over3}}\overline{u}\gamma_{\mu}u 
+ {\textstyle{{-1\,}\over{\ 3}}}\overline{d}\gamma_{\mu}d)
+ {\textstyle{{-1\,}\over{\ 3}}}\overline{s}\gamma_{\mu}s)
| p (n) \rangle
\label{eq:jmu}
\end{equation}
is the hadronic matrix element of the electromagnetic current operators for the proton (neutron).

The corresponding nucleon form factors for the virtual photon have three contributions:
\begin{equation}
G^{\gamma}_{\rm^{E,M}} = {\textstyle{2\over3}}G^u_{_{E,M}} 
+ {\textstyle{{-1\,}\over{\ 3}}} G^d_{_{E,M}} 
+ {\textstyle{{-1\,}\over{\ 3}}} G^s_{_{E,M}} 
\label{eq:GEM}
\end{equation}

The Z boson exchange between an electron and a nucleon leads to a similar structure of the current.
The contribution of $G^{Z}_{_{E,M}}$ could be observed thanks to the significant interference term 
in the matrix element of the scattering.
The measurement of the asymmetry of the longitudinally polarized electrons scattering from a proton
(left vs. right) allows us to find $G^s_{_{E,M}}$, see Refs.~\cite{DB2001,BB2005}
and complete flavor decomposition of the nucleon form factors (assuming isospin symmetry).

It is easy to see that the uncertainty in $G^s_{_{E,M}}$ (and $F_{1,2}$) contributes linearly to 
the uncertainty of $u$- and $d$-quark contributions. 
At \qsq=3.4 \gevsq~for the $\Delta G^s =$\GD~the corresponding uncertainty $\Delta (Q^4$\Ftwod$)\sim 0.35$, 
which is much larger than the contribution from the uncertainty of the \GEn~\cite{SR2010}, see Fig.~\ref{fig:F12ud}.

The interest in high \qsq~measurement of \Fsp~is also motivated by the expectation that \Fsp~has a maximum
at a momentum transfer much larger than the location of the \GEn~maximum due to the heavier mass of the $s$-quark.
Such an expectation is supported by the small radius of a $\phi$ meson which could be obtained
from the form factor in the $\phi$ decay to $\pi^\circ e^+e^-$~\cite{AA2016}.

There are two experimental difficulties in doing the \Fsp~measurement at large \qsq: 
the reduced counting rate and large background from inelastic electron-proton scattering.
The reduction of the counting rate, which is due to reduction of the $\sigma_{_{Mott}}G^2_{_{Dipole}}$, 
is partly compensated for by a linear increase of the asymmetry for high \qsq.
For suppression of the inelastic events we proposed to use the tight time and the angular correlations
between the scattered electron and recoiled proton (as well as the energy deposited in the detectors), 
as it was considered in Ref.~\cite{BW2005}.

\begin{figure}[ht]
\unitlength 1cm
\begin{minipage}[th]{8cm}
   \includegraphics[width=1.0\textwidth] {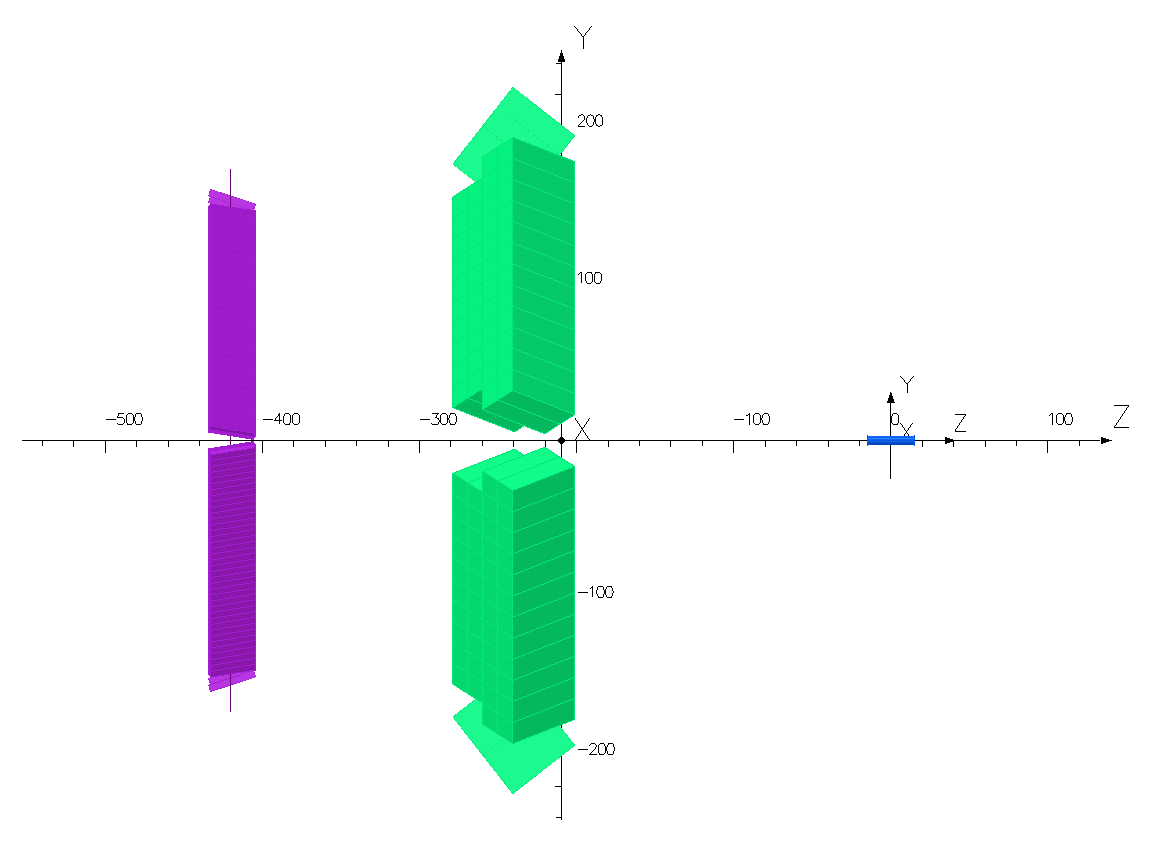} 
\end{minipage}
\begin{minipage}[th]{8cm}
\includegraphics[width=0.9\textwidth] {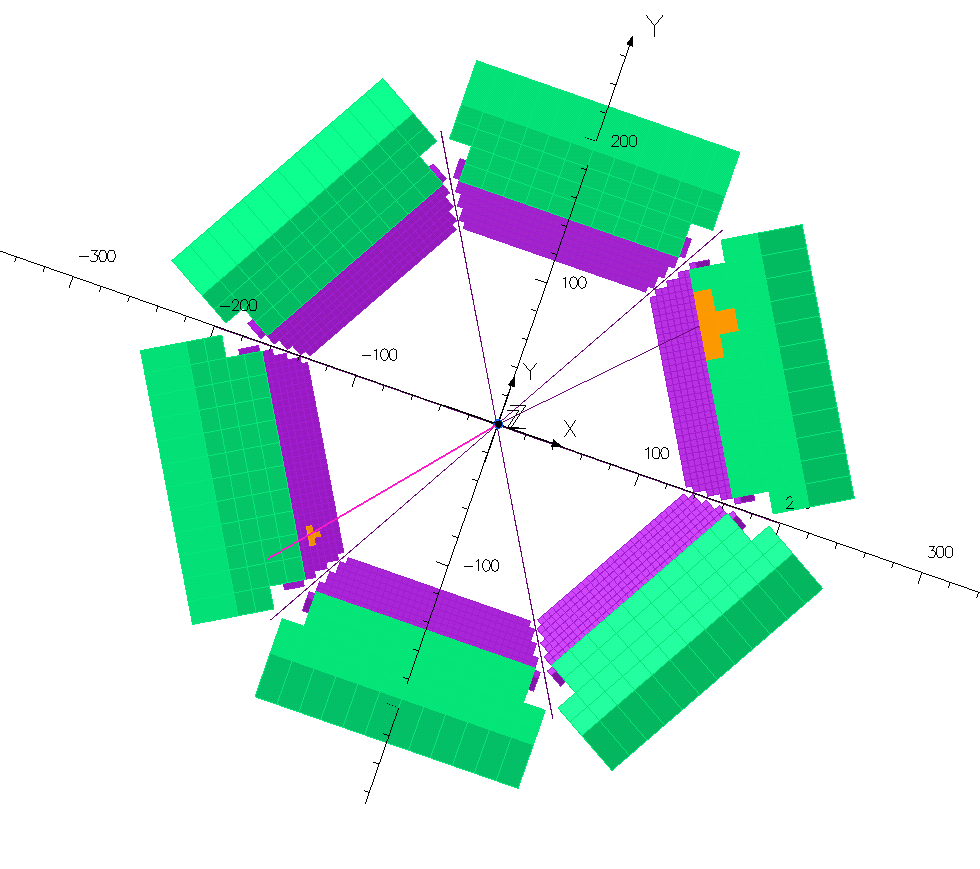}
\end{minipage}
   \caption{ Left: Side view of the apparatus. The electron beam goes from right to left.
   The proton detector is shown in green and the electron detector in purple; the liquid hydrogen
   target is shown in blue. Right: Front view of the apparatus. The blocks in orange get signals
   from the electron and the proton whose directions are shown in red.} 
   \label{fig:COP2}
\end{figure}

The solid angle of the apparatus should cover a suitable range of the momentum transfer $\Delta Q^2/Q^2 \sim 0.1$
for which the event rate variation over the acceptance is limited (we selected a factor of 4).
The equipment needed for such an experiment could be obtained from the SBS where a highly segmented
hadron calorimeter and electromagnetic calorimeters are under preparation for the GEp experiment~\cite{GEpE}.
Figure~\ref{fig:COP2} shows the proposed configuration of the detectors.
 
The proposed detector configuration has an electron arm with a solid angle of 0.1 sr at a scattering angle of 
18$\pm$1.5 degrees.
Within 30 days of data taking with a 6.6 GeV beam the PV asymmetry will be measured to 3\% relative accuracy which corresponds to an uncertainty of \Fsp~of 0.002.  
Such a measurement will provide the first experimental limit on \Fsp~at large momentum transfer 
of 3~\gevsq~(or discover its non-zero value) and reduce the current uncertainty from the strangeness contribution 
in the flavor separated proton form factors such as \Ftwod~by six times.

\section*{Acknowledgments}

This work was supported by Department of Energy (DOE) contract number DE-AC05-06OR23177, 
under which the Jefferson Science Associates operates the Thomas Jefferson National Accelerator Facility.

\end{document}